\documentclass[aps,prb,twocolumn,showpacs]{revtex4}
\usepackage{graphicx}
\begin{document}
%=======================================================================
%Zn$_{1-x}$Co$_x$O
%Zn$_{0.9}$Co$_{0.1}$O
%
\title{Electronic structures of Zn$_{1-x}$Co$_x$O using photoemission \\
	and x-ray absorption spectroscopy}

\author {S. C. Wi, J.-S. Kang$^{*}$, J. H. Kim, S.-B. Cho, B. J. Kim, 
	S. Yoon, and B. J. Suh} 

\affiliation{Department of Physics, The Catholic University of Korea,
        Puchon 420-743, Korea}

\author {S. W. Han and K. H. Kim}

\affiliation{Department of Physics, Gyeongsang National University,
        Chinju 660-701, Korea}

\author {K. J. Kim, B. S. Kim, H. J. Song, and H. J. Shin}

\affiliation{Pohang Accelerator Laboratory (PAL), 
	POSTECH, Pohang 790-784, Korea}

\author {J. H. Shim and B. I. Min}

\affiliation{Department of Physics, POSTECH, Pohang 790-784, Korea}

\date{\today}

\begin{abstract}

Electronic structures of Zn$_{1-x}$Co$_x$O have been investigated 
using photoemission spectroscopy (PES) and x-ray absorption
spectroscopy (XAS). 
The Co $3d$ states are found to lie near the top of the O $2p$ 
valence band, with a peak around $\sim 3$ eV binding energy.
The Co $2p$ XAS spectrum provides evidence that the Co ions 
in Zn$_{1-x}$Co$_{x}$O are in the divalent Co$^{2+}$ ($d^7$) states 
under the tetrahedral symmetry. 
Our finding indicates that the properly substituted Co ions
for Zn sites will not produce the diluted ferromagnetic semiconductor
property.

\end{abstract} 

%\pacs{79.60.-i, 75.47.Lx, 71.30.+h}

\maketitle

%\narrowtext
%==============================================================================
%\section{Introduction}
%\label{sec:intro}
   
%Spintronics, which utilizes both the spin and charge degrees of 
%freedom, becomes an important field in magnetism. 
Diluted magnetic semiconductors (DMSs) are considered as good 
candidates for spintronics materials because of the possibility of 
incorporating the spin degree of freedom in traditional semiconductors 
\cite{Furd88,Wolf01}.
Motivated by the recent theoretical calculations by Dietl {\it et al.} 
\cite{Dietl00} which predicted room-temperature ferromagnetism in 
Mn-doped ZnO, Zn$_{1-x}$T$_x$O systems (T$=3d$ transition-metal atom)
have been investigated extensively.  
Indeed there were reports that Zn$_{1-x}$Mn$_x$O epitaxial thin films 
exhibited ferromagnetic properties \cite{Fuku99,Jung02} and that
Zn$_{1-x}$Co$_x$O films showed ferromagnetism
with the Curie temperature $\rm T_C \sim 300$ K \cite{Ueda01}.
However, the reproducibility is somewhat questionable,
and there are contradicting reports on Co-doped ZnO
\cite{Ueda01,Kim02,Lee02}.
Technologically ZnO-based DMSs of Zn$_{1-x}$T$_x$O attract much 
attention due to the cheapness and abundance of ZnO,
and the easy solubility of T atoms in ZnO
up to several ten $\%$ \cite{Fuku99,Jin00}.  
In addition, since ZnO has a wide band gap energy of $\sim 3.4$ eV,
ZnO-based DMSs would be useful for short wavelength magneto-optical 
applications.

%In spite of extensive studies on the magnetic properties of 
%Zn$_{1-x}$T$_x$O-type DMSs, the origin of the observed ferromagnetism 
%in Zn$_{1-x}$T$_x$O has not been clarified yet.
In order to understand the magnetic interaction in Zn$_{1-x}$T$_x$O,
it is essential to understand the electronic structure of the doped 
$3d$ T impurities in Zn$_{1-x}$T$_x$O.
%Two competing factors are expected, such as 
%the strong on-site Coulomb interaction between $3d$ electrons 
%($U_{dd}$), and the hybridization between $3d$ and the host 
%valence-band electrons ($V_{pd}$) \cite{Mizo02}.   
Photoemission spectroscopy (PES) is one of the powerful experimental
methods for providing direct information on the electronic structures
of solids.
To our knowledge, 
only a few PES studies have been reported on Zn$_{1-x}$T$_x$O:
the PES study on Zn$_{1-x}$Mn$_x$O \cite{Mizo02}
and the Co $2p$ core-level PES spectra for Zn$_{1-x}$Co$_x$O
\cite{Lee02}.

In this study we have investigated the electronic structure of 
homogeneous bulk samples of Zn$_{1-x}$Co$_x$O using
PES and x-ray absorption spectroscopy (XAS).
Differently from previously reported film samples\cite{Ueda01}, 
magnetic properties of our bulk Zn$_{1-x}$Co$_x$O samples revealed
that the Co-Co magnetic
interaction is dominated by the antiferromagnetic coupling
\cite{Yoon03}.
Magnetic susceptibility at high temperature exhibits a typical Curie-Weiss
behavior with negative $\rm T_C$ and the magnetization of Co ion
is reduced with increasing the Co ion concentration reflecting
an increase in average antiferromagnetic interaction between Co ions.
Hence the present bulk sample is not a diluted ferromagnetic semiconductor.

%\section{Experimental and Calculational Details}
%\label{sec:exp}

Polycrystalline Zn$_{1-x}$Co$_x$O samples ($x=0, 0.1$) were synthesized
using the standard solid-state reaction method.
Zn$_{1-x}$Co$_x$O samples were obtained by sintering a mixed powder 
of ($1-x$)ZnO$+x$CoO at T $\simeq 1400 ^{\circ}$ C for $2$ hours
followed by the furnace cooling in air.
The x-ray diffraction (XRD) analysis showed that all the samples have 
the single phase of the Wurtzite structure with no impurity phase.
Valence-band PES, XAS and Co $2p \rightarrow 3d$ resonant photoemission 
spectroscopy (RPES) measurements were performed at 
the 2B1 and 8A1 beamlines of the PAL.
Samples were cleaned {\it in situ} by repeated scraping with a diamond 
file and the data were obtained at room temperature with the pressure 
better than $4 \times 10^{-10}$ Torr.
The Fermi level $\rm E_F$ \cite{calib} and the overall instrumental 
resolution (FWHM) of the system were determined from the valence-band 
spectrum of a sputtered Au foil in electrical contact with a sample.
The FWHM was about $0.1-0.6$ eV between a photon energy $h\nu\sim 30$
eV and $h\nu\approx 600$ eV.
All the spectra were normalized to the mesh current.
The XAS spectra were obtained by employing the total 
electron yield method.
The experimental energy resolution for the XAS data was set
to $\sim 0.2$ eV at the Co $2p$ absorption threshold
($h\nu \approx$ 700 eV).
Core-level spectra and the valence-band spectrum with $h\nu=1486.6$
eV were obtained by using the monochromatized Al $K\alpha$ radiation
with a FWHM of $\sim 0.6$ eV.
%For both PES and XAS measurements, samples were cleaned 
%{\it in situ} by repeated scraping with a diamond file.

%\section{Results and Discussion}
%\label{sec:results}

%-------------------------------------
\begin{figure}[t]
\includegraphics[scale = 0.5]{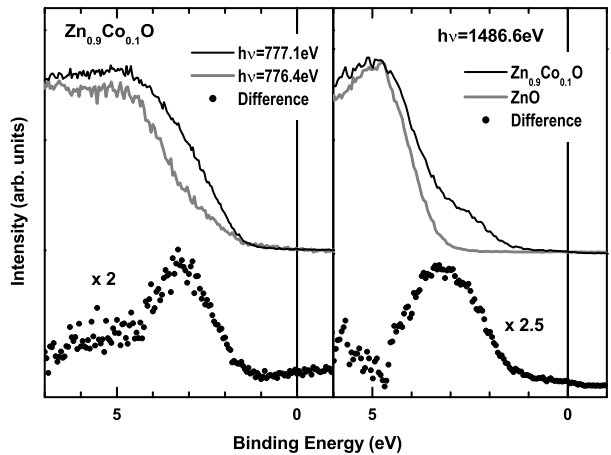}
\caption{Left: Comparison of the Co $2p_{3/2} \rightarrow 3d$
	on-resonance spectrum ($h\nu
	\approx 777.1$ eV) and the off-resonance spectrum 
	($h\nu\approx 776.4$ eV) for Zn$_{0.9}$Co$_{0.1}$O,
	scaled at $\sim 7$ eV BE. 
	Right: Comparison of the valence-band spectrum of ZnO
	and that of Zn$_{0.9}$Co$_{0.1}$O, obtained with
	$h\nu=1486.6$ eV. 
	For both cases, the difference curves are shown at the bottom.}
\label{3d} 
\end{figure}
%-------------------------------------
\begin{figure}[t]
\includegraphics[scale = 0.65]{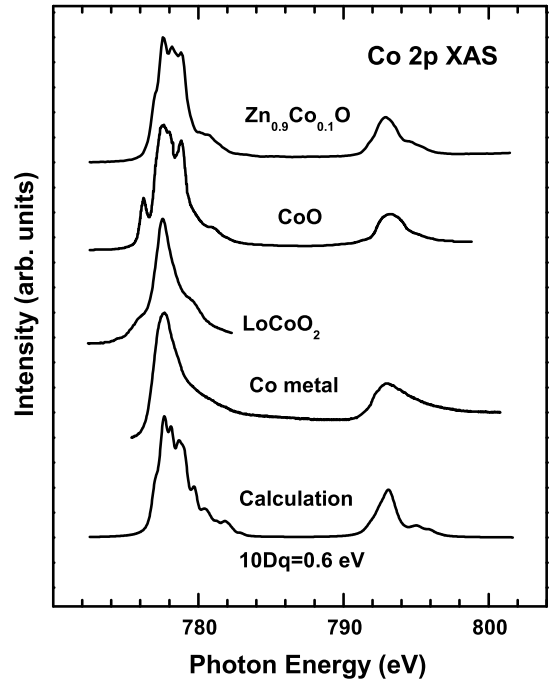}
\caption{
	The Co $2p$ XAS spectrum of Zn$_{0.9}$Co$_{0.1}$O 
	in comparison to those of 
	CoO (Co$^{2+}$) [Ref.~\cite{Regan01}], 
	LiCoO$_2$ (Co$^{3+}$) [Ref.~\cite{Elp91}],
	and Co metal [Ref.~\cite{Regan01}]. 
	The data for CoO, LiCoO$_2$, and Co metal were shifted 
	by $+0.1$ eV, $-3.64$ eV and $-0.1$ eV, respectively. 
	The calculated Co $2p$ XAS spectrum
	for low-spin Co$^{2+}$ under the $T_d$ symmetry is presented
	at the bottom.  } 
\label{2pxas} 
\end{figure}
%-------------------------------------

We have measured the valence-band PES spectra of
Zn$_{1-x}$Co$_x$O ($x=0,~0.1$) over a wide several photon energy
range 25 eV $\le h\nu \le$ 1486.6 eV, which includes
the Co $2p \rightarrow 3d$ RPES energies (see Fig.~\ref{2pxas}).
Our valence-band PES spectrum of ZnO agrees with the existing
data in the literature \cite{Ruckh94}.
Valence-band spectra for ZnO and Zn$_{0.9}$Co$_{0.1}$O 
were rather similar, showing two large structures, one sharp peak 
around 11 eV binding energy (BE), 
and the other broad band between $\sim 3$ eV and $\sim 7$ eV BE.
The former and latter peaks are assigned as the Zn $3d$
and O $2p$ emissions, respectively. 
They also reveal negligible emission near $\rm E_F$, reflecting 
that both ZnO and Zn$_{0.9}$Co$_{0.1}$O are insulating.

The left panel of Fig.~\ref{3d} compares 
the on-resonance ($h\nu \approx 777.1$ eV) and off-resonance 
($h\nu\approx 776.4$ eV) spectra for Zn$_{0.9}$Co$_{0.1}$O 
in the Co $2p_{3/2} \rightarrow 3d$ RPES, 
which are scaled to each other at $\sim 7$ eV BE.
The difference between on-resonance and off-resonance 
can be considered as representing the Co $3d$ partial spectral 
weight distribution (PSW)
\cite{Kang02}.
The right panel of Fig.~\ref{3d} compares the valence-band spectrum 
of Zn$_{0.9}$Co$_{0.1}$O to that of ZnO, obtained with $h\nu=1486.6$ eV.
The differences between
$x=0$ and $x=0.1$ also reflect the effect of the Co $3d$ states. 
In both cases the differences are shown at the bottom, which 
are very similar to each other, showing a peak around $\sim 3$ eV BE 
and negligible emission near $\rm E_F$ \cite{scale}.
These results provide evidence that the Co $3d$ states in 
Zn$_{1-x}$Co$_x$O are located near the top of the O $2p$ valence band. 

The extracted Co $3d$ PSW reveals almost no emission between $\rm E_F$ 
and 2 eV BE. 
This feature does not agree with the calculated 
Co $3d$ PDOS (partial density of states), 
obtained from the LSDA (local spin-density approximation) band 
structure calculation \cite{Sato02}. 
The calculated Co $3d$ PDOS for Zn$_{0.75}$Co$_{0.25}$O
shows a sharp $e_{g}^2\downarrow$ peak near $\rm E_F$. 
To resolve this discrepancy,
more elaborate calculations such as LSDA$+U$ incorporating
the Coulomb correlation $U$ between Co $3d$ electrons 
will be required because it is well known that the LSDA 
calculation does not properly describe the electronic structures 
of insulators with strongly correlated electrons \cite{Park00}.

Figure~\ref{2pxas} compares the Co $2p$ XAS spectrum of 
Zn$_{0.9}$Co$_{0.1}$O with those of reference Co oxides, having 
the formal Co valences of $2+$ (CoO, Ref.~\cite{Regan01}) 
and $3+$ (LiCoO$_2$, Ref.~\cite{Elp91}),
and that of Co metal (Ref.~\cite{Regan01}).
Note that Co ions both in CoO and LiCoO$_2$ are located at 
the octahedron ($O_h$) centers.
The peak positions and the line shape of the T $2p$ XAS spectrum 
depend on the local electronic structure of the T ion, 
providing the information about the valence state 
and the ground state symmetry of the T ion \cite{Groot90,Laan92,Kim03}.
The Co $2p_{3/2}$ and $2p_{1/2}$ spectral parts are clearly separated 
by the large $2p$ core-hole spin-orbit interaction, and the core-hole 
lifetime broadening is small, resulting in the sharp multiplet 
structures.
The Co $2p$ XAS spectrum of Zn$_{0.9}$Co$_{0.1}$O looks similar to 
that of CoO except for the absence of the low-$h\nu$ shoulder 
($h\nu\sim 776$ eV), but quite different from those of LiCoO$_2$ and Co metal.
This observation
indicates that Co ions in Zn$_{1-x}$Co$_x$O might be
in the divalent Co$^{2+}$ valence states, but not in the trivalent 
Co$^{3+}$ valence states, and that the formation of the Co metal 
cluster in our Zn$_{1-x}$Co$_x$O samples can be ruled out. 

In order to estimate the valence state of Co ions in the ground states,
we have analyzed the Co $2p$ XAS spectrum of Zn$_{0.9}$Co$_{0.1}$O 
within the cluster model where the effects of the multiplet 
interaction, the crystal field, and the hybridization 
with the O $p$ ligands are included \cite{Laan92}.
The calculated XAS spectrum is shown at the bottom of Fig.~\ref{2pxas}. 
We have found that including only one configuration, corresponding to 
Co$^{2+}$ ($d^7$), and the tetrahedral ($T_d$) crystal field energy
of $10Dq=0.6$ eV yields
a good fit for the measured $2p$ XAS spectrum \cite{cluster}. 
The analysis shows that it is not necessary to include 
the charge-transfer configuration, $d^{n+1}\underbar{L}^{1}$ 
($\underbar{L}^{}$: a ligand hole), indicating that the hybridization 
between Co $3d$ and O $p$ orbitals is small, in contrast 
to Mn-doped ZnO case \cite{Mizo02}.
We have checked that small changes in the parameters do not affect
the overall spectral shape including the prominent peak positions.
Based on the calculated XAS spectrum, one can conclude that 
the doped-Co ions in Zn$_{1-x}$Co$_{x}$O are divalent in the ground 
states, i.e., Co$^{2+}$ ($S=3$) 
under the $T_d$ symmetry with a small crystal field energy \cite{LS}. 
Our finding suggests that, in our samples, 
Co ions substitute properly 
for the Zn sites with Co$^{2+}$ valence states. In view of no long 
range ferromagnetic order in our samples, 
the reported DMS properties 
in some Co-doped ZnO \cite{Ueda01,Lee02} are likely to be extrinsic, 
that is, to originate from Co cluster or other impurity phases.

%\section{Conclusions}
%\label{sec:conc}

In conclusion, the electronic structures of bulk Zn$_{1-x}$Co$_x$O 
samples have been investigated by employing PES and XAS. 
According to the Co $2p \rightarrow 3d$ RPES and the careful comparison
between $x=0$ and $x=0.1$, the Co $3d$ states in Zn$_{1-x}$Co$_x$O are
found to lie near the top of the O $2p$ valence band with a peak
around $\sim 3$ eV BE. The measured Co $2p$ XAS spectrum shows that  
the ground states of Co ions in Zn$_{1-x}$Co$_{x}$O are the divalent
Co$^{2+}$ states under the $T_d$ symmetry corresponding to the total 
spin of $S=3$ per Co ion.
Our finding suggests that the properly substituted Co ions
in Co-doped ZnO do not produce the DMS property with the long range 
ferromagnetic order.

Acknowledgments$-$
This work was supported by the KRF (KRF-2002-070-C00038) and 
by the KOSEF through the CSCMR at SNU and the eSSC at POSTECH.
The PAL is supported by the MOST and POSCO in Korea.


\begin{thebibliography}{99}
\bibitem{calib} Since Zn$_{1-x}$Co$_x$O samples are insulating,
        there is some ambiguity in determining $\rm E_F$. 
	However, the charging did not distort the valence-band 
	lineshape with time.
\bibitem{scale} When the difference is obtained with the 
	on-resonance and off-resonance spectra being scaled at 
	the Zn $3d$ peak, we get another broad feature between
	$\sim 4$ eV and $\sim 8$ eV BE. This feature is ascribed
	to the Co $3d$-O $2p$ hybridized states.
	Here we focus on the resonating feature which has rather 
	pure Co $3d$ electron character.
\bibitem{cluster} 
	The fitting parameters are scaled by 80\% of the atomic 
	Slater integrals ($F(dd),F(pd)$, and $G(pd)$).
\bibitem{LS} In the $T_d$ symmetry for Co ions, 
	one cannot distinguish between the low-spin (LS) and 
	high-spin (HS)
	divalent Co$^{2+}$ states because the $e_g$ states are located 
	below the $t_{2g}$ states.
	This results in the same total spin $S=3$ for both 
	the small (the so-called LS state) and 
	high (the so-called HS state) values of $10Dq$.
\end{thebibliography}
\end{document}